\documentclass[11pt]{article}
\usepackage{hyperref}
\pdfoutput=1

\begin{document}
\title{The turbulent/nonturbulent interface in boundary layers}
\author{G. Borrell, J. A. Sillero and Javier Jim\'enez \\
\vspace{6pt} School of Aeronautics,\\
Universidad Polit\'ecnica de Madrid}

\maketitle

\begin{abstract}
  This video shows three-dimensional contours at two different values
  of enstrophy for a turbulent boundary layer up to $Re_\theta \simeq
  2500$. Its intention is to show how, what can be be considered the
  turbulent/nonturbulent
  interface for a turbulent boundary layer, evolves in time.\\
  \\
  \vspace{6pt} Keyword: fluid dynamics video
\end{abstract}

\section{Introduction}

The flow simulated is a turbulent boundary layer from $Re_\theta=300$
to $\simeq 2500$, where only the last fourth of the computational
domain has been represented. The portion shown is $10\delta_{99}$ wide
and $12\delta_{99}$ long. The visualized width corresponds with the
simulated width, hence the contours are periodic in the spanwise
direction. All the contours presented are coloured with the distance
to the wall.

The first scene corresponds to a contour for a value of enstrophy,
defined as the magnitude of the vorticity fluctuations, equal to 10\%
the value at the wall. This value is one order of magnitude higher to
what is usually taken as a threshold for the turbulent / nonturbulent
interface, and tube-like structures are clearly seen in the region
further from the wall. The same temporal evolution is repeated several
times.

In the second scene the previous threshold is changed from 10\% the
value of the wall to 1\%, a value that generates a connected and
relatively smooth surface. This second value for the threshold is in
the range of what is used to research the turbulent / nonturbulent
interface as a surface.

The third scene is the same temporal evolution of the first one, but
with the contour taken for the second value of enstrophy.
 
The video can be downloaded in different resolutions from the research
group's website: 
\url{http://torroja.dmt.upm.es}

\end{document}